      [1999/12/01 v1.4c Il Nuovo Cimento]
\documentclass{cimento}
\usepackage{epsfig}
\title{Search for Gamma Ray Bursts at Chacaltaya} 
\author{A.~Castellina\from{ins:1},
P.L.~Ghia\from{ins:1},
F.~Kakimoto\from{ins:3},\ETC
T.~Kaneko\from{ins:4},
C.~Morello\from{ins:1},\\
G.~Navarra\from{ins:2},
K.~Nishi\from{ins:5},
O.~Saavedra\from{ins:2},
G.~Trinchero\from{ins:1},
D.~Urzagasti\from{ins:6},\\
P.~Vallania\from{ins:1},
A.~Velarde\from{ins:6},
\underline{S.~Vernetto}\from{ins:1} \atque
H.~Yoshii\from{ins:7}}
\instlist{
\inst{ins:1} Istituto di Cosmogeofisica del CNR, Torino, Italy
\inst{ins:2} Dipartimento di Fisica Generale dell'Universita' di Torino, Italy
\inst{ins:3}Department of Physics, Tokyo Institute of Technology, Meguro,
Tokyo 152, Japan
\inst{ins:4}Department of Physics, Okayama University, Okayama 700, Japan
\inst{ins:5} Institute of Physical and Chemical Research, Wako, Saitama
351-01, Japan
\inst{ins:6}Instituto de Investigaciones Fisicas, Universidad Mayor de San
Andres, La Paz, Bolivia
\inst{ins:7}Department of Physics, Ehime University, Ehime 790, Japan
}
\PACSes{\PACSit{98.70Rz}{Gamma Ray Bursts}}
\begin{document}

\maketitle

\begin{abstract}
A search for Gamma Ray Bursts in the GeV-TeV
energy range has been performed
by INCA, an air shower array working
at 5200 m of altitude at the Chacaltaya Laboratory (Bolivia).
The altitude of the detector and the use  of the
``single particle technique''  allows to 
lower the energy threshold up to few GeVs. 
No significant signals are observed during the occurrences
of 125 GRBs detected by BATSE, and the obtained upper limits
on the energy fluence in the interval 1-10$^{3}$ (1-10$^2$) GeV,
range from 3.2 (8.6) 10$^{-5}$ to 2.6 (7.0) 10$^{-2}$ erg cm$^{-2}$ 
depending on the zenith angle of the events.
These limits, thanks to the extreme altitude of INCA,
are the lowest ever obtained in the sub-TeV energy region
by a ground based experiment.    

\end{abstract}

\section{High energy gamma rays from GRBs}

Since their discovery in 1973 most of our knowledge on Gamma Ray Bursts 
(GRBs) has been confined to the KeV-MeV energy region; only during the 
last 3 years the observations extended to other wavebands, namely
X, optical, infrared and radio, 
yielding a lot of exciting results and producing a 
break-through in our understanding of these misterious events 
(for a review see~\cite{ref:gal} and~\cite{ref:klo}).
Now we know that GRBs are huge explosions occurring in 
far galaxies and are
probably the most energetic events in the universe, exceeding
even Supernovae. The cause of the
release of such amount of energy is still unknown and what we observe 
is probably only the effect of relativistic shock waves 
produced by the explosion interacting each other
or with the surrounding interstellar medium. 
Even if the ``central engine'' remains
hidden the study of the radiation emitted in all wave bands
by the interacting shock waves gives important informations on 
the acceleration mechanisms
and on the ambient medium in which the shock  propagates
(see~\cite{ref:pir} for a theoretical review).

So far the radiation emitted by GRBs in the GeV-TeV band 
has been very poorly studied
due to the extremely low fluxes.
EGRET, the high energy
detector aboard the satellite GRO, during its life 
detected only few very intense
events containing some GeV photons~\cite{ref:egr}.  
However all GRBs could contain
a GeV energy component, so far not measured 
only because of the low flux. 
In fact, many
models~\cite{ref:mes,ref:vie,ref:bar,ref:der,ref:pao,ref:tot}
predicts emissions in the GeV-TeV region. 
Upper limits in the 10-100 TeV region has been obtained by several
air shower arrays ~\cite{ref:eas,ref:cyg,ref:heg,ref:tib}.
An interesting TeV candidate 
has been observed by the ground based Milagrito 
detector in coincidence with GRB970417~\cite{ref:mil}.

The paucity of the flux is not the only problem to face when studing
the high energy component of the GRB spectrum. The major problem
(and unsolvable) is the absorption of gamma rays in the intergalactic
space. GeV and TeV gamma rays interact with the infrared photons
emitted by stars and dust producing electrons and positrons pairs.
The flux of photons of energy $E$ decreases as 
$dN/dE = (dN_0/dE) e^{-\tau(E,z)}$, where $z$ is the redshift of the
source. The optical depth $\tau$ increases with
$E$ and $z$ but it is not easy to evaluate, due to the
difficulty of measuring the infrared field in the far universe.
According to a model~\cite{ref:ste} 
the optical depth becomes 
equal to 1 for energies as low as $E \sim$  40-70 (200-400) GeV when
$z$=1(0.2).
 
So far about 15 redshifts of GRBs hosts have been measured: they range
between 0.4 and 4.5, clustering around $z \sim$ 1.
This implies that even if GRBs emit TeV-PeV gamma-rays 
we unlikely could detect them, unless we are so lucky to observe
a rare event occuring very close to us.
As a consequence, to study the high energy component of GRBs 
we are forced to concentrate our efforts in
the region of energy less than 1 TeV.
    
Ground based experiments can study high energy GRBs 
detecting the secondary particles of air showers
generated by the interactions of gamma
rays with the atmosphere nuclei. They can be divided
in two major classes: Cerenkov Telescopes, detecting the Cerenkov photons
emitted by the electrons and positrons traveling throught the atmosphere, 
and Air Shower Arrays, detecting the charge particles (mainly electrons
and positrons) that reach the ground. Cerenkov telescopes are not
suitable to detect transient events as GRBs, because of their small
field of view (few squared degrees) and their limited duty cycle ($\sim$
10$\%$). On the contrary, Air Shower Arrays have a larger field of view
($\sim \pi$ sr) and a duty cycle of $\sim$ 100$\%$. 
They usually work with energy thresholds of $\sim$ 10-100 TeV,
but in searching for transient events as GRBs, they can lower
the threshold to few GeVs using the ``single particle technique''
and operating at very high altitude.

The single particle technique~\cite{ref:ver,ref:eas}
consists in counting all the particles giving a signal
in the detector (not requiring any coincidence between particles
as it is usually done to detect air showers). In this way it is possible
to detect the lonely survivals of small 
showers produced by primaries of relatively low energy. Obviously with
only one particle per shower it is not
possible to reconstruct the arrival direction nor the energy of the
primary and a GRBs is detectable only as a short duration excess over the
single particle background, possibly in coincidence with a GRB
satellite detection.
The background is due to all secondary cosmic rays from all the
sky above the horizon.

The signal from a GRB strongly increases with the altitude of the
detector ~\cite{ref:ver}. 
Fig.1 shows the mean number of particles reaching the ground
generated by a gamma ray of different energies as a function of the
altitude (the values refer to a zenith angle $\theta$=30$^{\circ}$). 
Going from 2000 to 5000 m, a 10 (100) GeV signal increases by 
a factor $f_s \sim$100 (50); since the background increases only by a factor
$f_b \sim$ 4, the  sensitivity will improve by
a factor $f = f_s/\sqrt{f_b} \sim$ 50 (25).

Following these considerations we decided to exploit the air
shower array BASJE, operating since 1962 (with several
upgrades since then) at the Chacaltaya Laboratory
at 5200 m a.s.l., to develop the experiment INCA, 
to search for GeV GRBs using the single particle technique.

\begin{figure}[htb]
\vfill \begin{minipage}{.45\linewidth}
\begin{center}
\mbox{\epsfig{figure=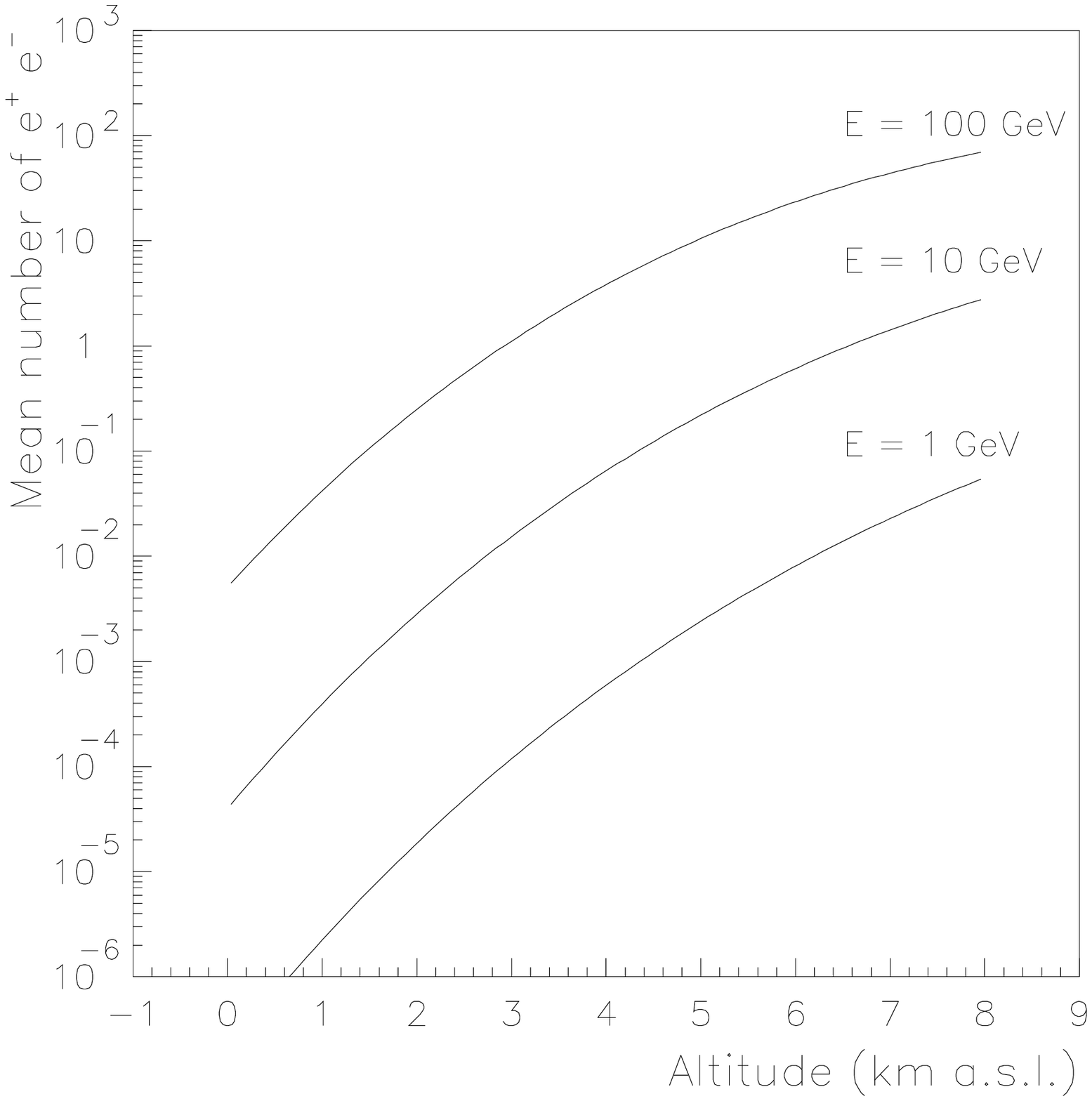,width=1.0\linewidth}}
 \end{center}
\vskip -0.5cm
\caption{Mean number of e$^{\pm}$ reaching the ground
generated by a gamma ray of different energies 
as a function of the altitude ($\theta$=30$^{\circ}$).}
\label{fi:np}
\end{minipage}\hfill
\begin{minipage}{.45\linewidth}
 \begin{center}
\mbox{\epsfig{figure=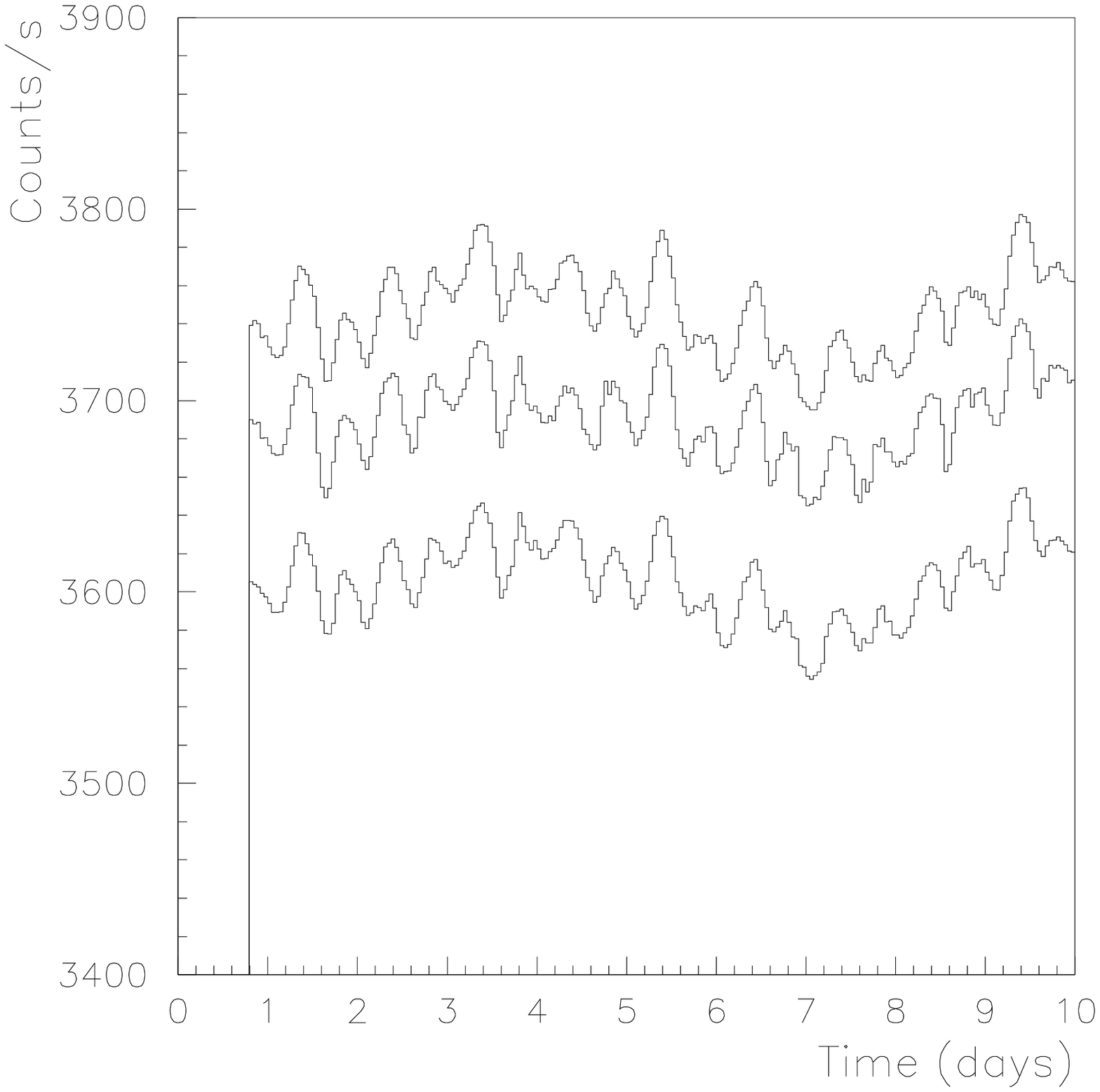,width=1.0\linewidth}}
 \end{center}
\vskip -0.5cm
\caption{Counting rate of 3 INCA scintillator detectors during $\sim$ 
10 days of measurement.}
\label{fi:osc}
\end{minipage}\vfill
\end{figure}

\section{The INCA experiment and its sensitivity}

The INCA experiment~\cite{ref:in1,ref:in2} consists of 
12 scintillator detectors of 2 x 2 m$^2$
area distributed over an area of $\sim$ 20 x 20 m$^2$.
Each detector is viewed by a photomultiplier of 15 cm diameter
and works with an energy threshold of $\sim$ 10 MeV.
The experiment is running since December 1996.
In August 2000 INCA has been upgraded with 16 muon detectors,
each consisting of a 4 m$^2$ scintillator operating under a depth
of 320 g cm$^{-2}$.
The muon detectors are used as a sort of ``anticoincidence''
to reject possible excesses in the counting rate not due to a gamma ray burst, 
since a true gamma ray signal is expected not to
contain muons.  
In the data presented in this paper the muon detectors are not yet
implemented. 

Every second the data acquisition records:

a) the number of counts of each detector

b) the atmospheric pressure

c) the universal time with a precision better than 1 $\mu$s.

The counting rate (mainly due to electrons, positrons and muons)
from low energy cosmic ray air showers is $C_d \sim$ 3700-3800 count s$^{-1}$
for each detector
and it is modulated by the atmospheric pressure, the 24 hours anisotropy
and the solar activity.

Fig.2 shows the behaviour of 3 detectors during $\sim$ 10 days of
measurement, where  variations
of amplitude $\sim 2-3 \%$ on time scales ranging from a fraction of a day
to several days are visible.
However it is important to note that
these modulations do not hamper the GRBs search because
the time scales of the phenomena are quite different.
More troublesome are electrical noises that can simulate short time
variations in the single particle flux. As a consequence it is 
very important to check accurately the behaviour of the detectors
in order to avoid the presence of spurious signals. 

The average INCA total background rate is $C_b \sim$ 4.5 10$^4$ s$^{-1}$.
As a consequence,  
a GRB of time duration $\Delta t$ = 1 s is detectable with a significance
of 4 standard deviations if the number of detected particles is larger
than $C_{th}$ = 4 $\sqrt{C_b} \sim$ 840, while for a generic $\Delta t$ = $t$
s, $C_{th}$ increases by a factor $\sqrt{t}$.
To have an idea of the number of particles expected from a GRB,
we considered 14 GRBs detected by EGRET whose spectrum has been
published~\cite{ref:egr}. All the detected spectra show a power law behaviour
without breaks up to the maximum energy that the EGRET sensitivity
could observe, with an average slope  $\alpha \sim$ -2.2.
For simplicity we have assumed that each spectra  extends with
a slope equal to the measured one 
up to a cutoff energy $E_{max}$ and than zeroes
(due to an intrinsic cutoff at the source or to the
intergalactic absorption).
We have calculated the number of particles
produced by the EGRET bursts for different values of $E_{max}$, assuming
a burst zenith angle $\theta$=0$^{\circ}$, 
and we have compared it with the
minimum number $C_{th}$ necessary to give a significant signal.
For $E_{max}$ = 1 TeV, 8 (4) bursts out of 14 are detectable if $\Delta t$ =
1 (10) s., while for $E_{max}$ = 100 GeV the number of detectable bursts 
is reduced to 4 (3).
These considerations show that INCA 
could detect GRBs of intensity comparable to
the most intense observed so far by EGRET, provided that the energy spectrum
extends at energy E $\ge$ 100 GeV and that they occure at small zenith angles.

\section{Data analysis and results}

The data analysis consists in the search for significant excesses
in the scintillator counting rates during the occurrence of the bursts
observed by the BATSE instrument aboard the CGRO satellite 
(orbiting since April 1991 to June 2000).
In the period December 1996 - March 2000 125 BATSE bursts have occurred
in the INCA field of view (i.e. zenith angle $\theta <$ 60$^{\circ}$).

For each BATSE event the INCA data recorded during 10000 s around
the burst time are selected and carefully analized to reject
possible spurious event due to the noisy behaviour of some detectors.
Finally the counts of each detector are summed and the total counting rate
distribution is studied to identify possible statistically
significant fluctuations.

We looked for excesses of different durations $\Delta t$ = 1, 2, 6, 10, 20,
50, 100, 200 s, inside a time interval of 2 hours around the burst time,
shifting the window position in step of $\Delta t$/2 (except the case
$\Delta t$= 1s, where the step is $\Delta t$).
The number of counts $C$ recorded in $\Delta t$ are compared with
the expected background $B$, calculated using the background rate during 30
minutes around $\Delta t$ (in 30 minutes the background modulations
are negligible).
We found no excess for any BATSE GRB and for any window $\Delta t$;
the obtained distributions of the fluctuations
$f = C - B$ follow the expectations of a uniform background.
Fig.3 shows the distribution of $f$ in unit of standard deviations
for four different time windows summed over 125 bursts. 
They are well fitted by  Gauss
distributions with r.m.s. ranging from 1.06 ($\Delta t$=1 s) to 1.29
($\Delta t$= 200 s).

\begin{figure}[htb]
\vfill \begin{minipage}{.45\linewidth}
\begin{center}
\mbox{\epsfig{figure=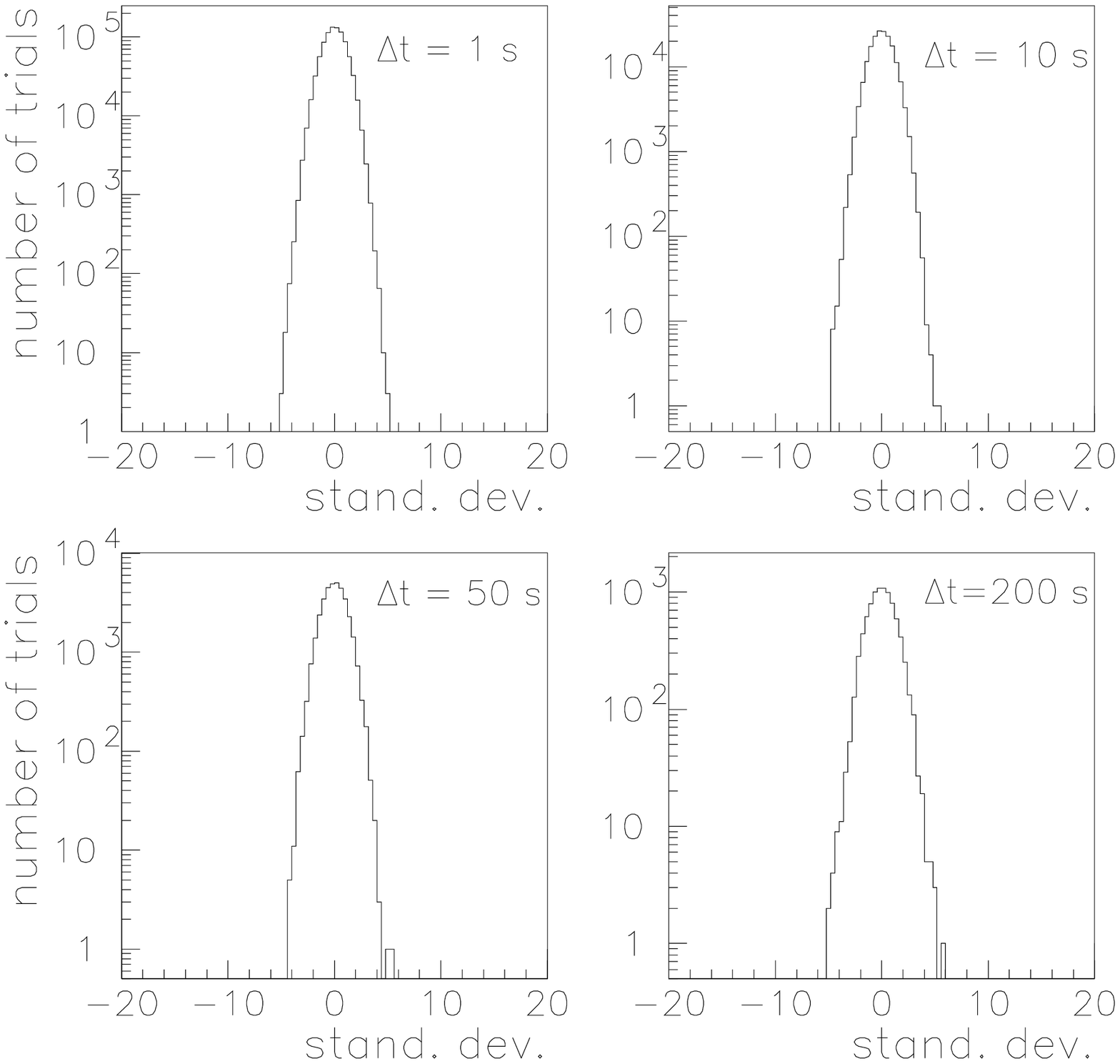,width=1.0\linewidth}}
 \end{center}
\vskip -0.5cm
\caption{Distribution of the fluctuations of counting rates in 2 hours
around the time occurences of 125 BATSE GRBs, for 4 different time windows.}
\label{fi:sig}
\end{minipage}\hfill
\begin{minipage}{.45\linewidth}
 \begin{center}
\mbox{\epsfig{figure=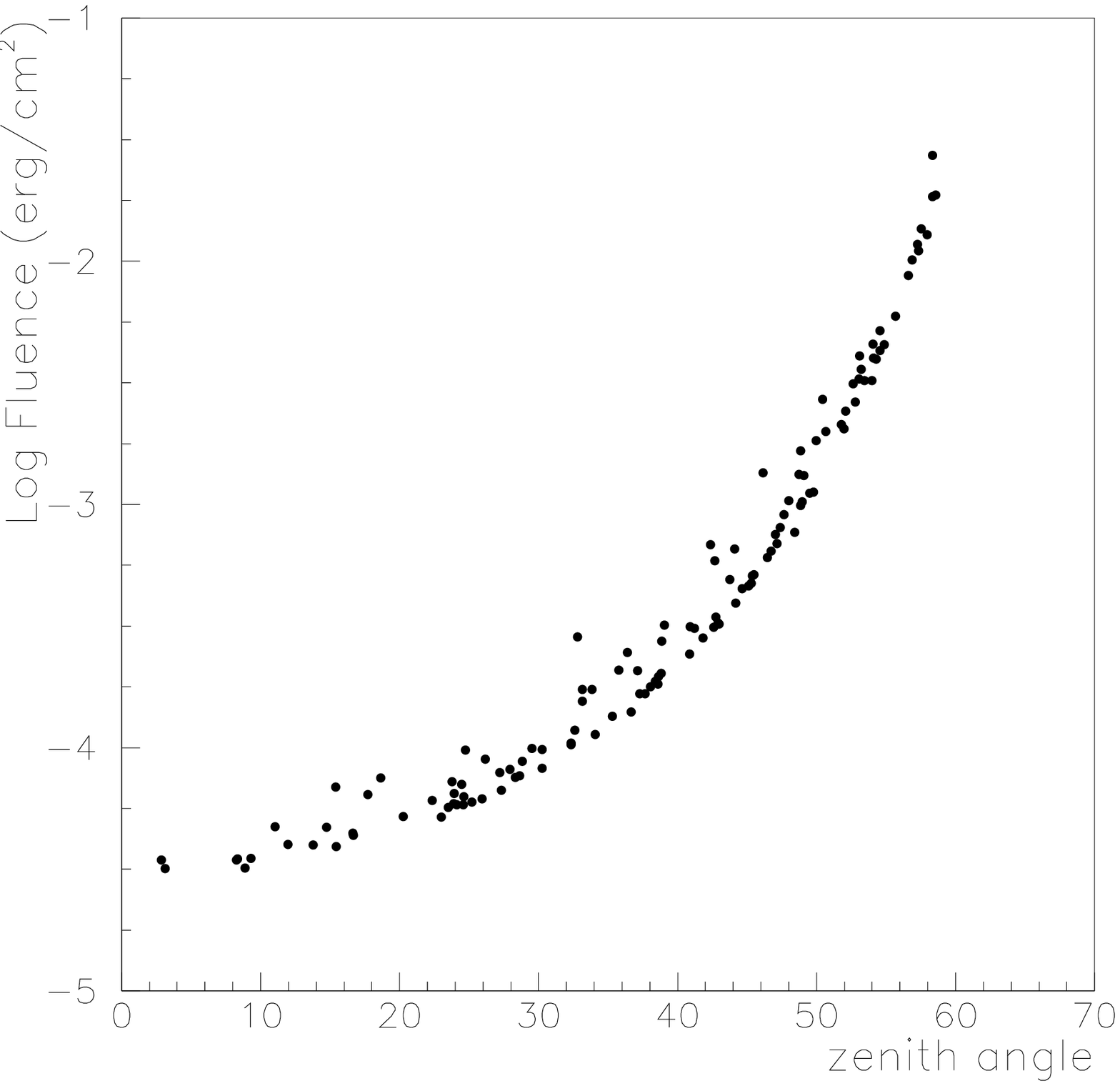,width=1.0\linewidth}}
 \end{center}
\vskip -0.5cm
\caption{Upper limits on the energy fluence in the 1 GeV-1 TeV energy interval
for 125 GRBs, in a time window $\Delta t$=10 s starting from the onset
of the burst, as a function of the zenith angle of the event.}
\label{fi:upl}
\end{minipage}\vfill
\end{figure}

Fig.4 shows the energy fluence upper limits in the energy range
1 GeV-1 TeV for the 125 bursts considered in this analysis, as a function
of the zenith angle of the event.
The fluences have been calculated at 4 standard deviations level
assuming the GRBs spectra as $dN/dE \propto E^{\alpha}$ with $\alpha$ = -2
extending up to 1 TeV and assuming a burst duration of 10 s.
They range from 3.2 10$^{-5}$ to 2.6 10$^{-2}$ erg cm$^{-2}$  
depending on the zenith angle of the event.
If the spectrum extends only up to 100 GeV (a more realistical assumption
due to the intergalactic absorption) 
the upper limits in the 1-100 GeV energy region
are a factor 2.7 higher than the previously given values.

A very simple technique as the
single particle detection, used by the highest air shower array 
operating in the world, has provided
the lowest upper limits on the GRBs 
flux ever obtained in the $\sim$ 1 GeV - 1 TeV energy region
by a ground based experiment.

\end{document}